\documentclass[runningheads]{llncs}
\usepackage{makeidx,fancybox}
\usepackage{amsmath,hhline,mathrsfs}
\usepackage{amssymb,array,amscd}
\usepackage[all,2cell]{xy}
\title{Supercompiling String Programs \\ Using Word Equations as Constraints}
\author{Antonina Nepeivoda \\
\institute{Program Systems Institute of Russian Academy of Sciences\thanks{The reported study was partially supported by RFBR, research project 
  No. 17-07-00285\_a, and Russian Academy of Sciences, research project No.~AAAA-A16-116021760039-0.}
\\
  Pereslavl-Zalessky, Russia\\}\email{a\_nevod@mail.ru}
}

\usepackage{color}            

\newcommand{\skipper}{\hspace{-7pt}\textcolor{white}{!_{a_a}}}

\newcommand{\ie}{{\it i.e.} }
\newcommand{\eg}{{\it e.g.}}
\newcommand{\etc}{{\it etc.} }

\newcommand{\st}{{\it s.t.} }
\newcommand{\via}{{\it via} }

\newcommand{\sectionname}{Sec.}

\newcommand{\Prog}{\mathsf{P}}
\newcommand{\Node}{\mathsf{N}}
\newcommand{\Rule}{\mathsf{R}}
\newcommand{\Ineq}{\mathit{I}\hspace{-0.5pt}}
\newcommand{\FPref}{\mathtt{Pref}}
\newcommand{\FEqual}{\mathtt{Equal}}
\newcommand{\FCheck}{\mathtt{Check}}
\newcommand{\FGram}{\mathtt{Gram}}

\newcommand{\Iter}{\mathtt{Iter}}
\newcommand{\Go}{\mathtt{Go}}

\newcommand{\BothA}{\mathtt{BothA}}
\newcommand{\Fun}{\mathtt{F}}
\def\FPal{\mathtt{Pal}}
\def\FGenA{\mathtt{RepA}}
\def\ConstrEq{\mathbf{E\hspace{-0.8pt}q\hspace{-0.8pt}u\hspace{-0.8pt}a\hspace{-0.8pt}l}}

\newcommand{\conc}{\,}

\newcommand{\cA}{\mathbf{A}}
\newcommand{\cB}{\mathbf{B}}
\newcommand{\cC}{\mathbf{C}}
\newcommand{\cI}{\mathbf{I}}
\newcommand{\cS}{\mathbf{S}}
\newcommand{\pw}{\mathit{w}}
\newcommand{\pu}{\mathit{u}}
\newcommand{\pv}{\mathit{v}}
\newcommand{\pt}{\mathit{s}}
\newcommand{\vx}{\mathtt{x}}
\newcommand{\vy}{\mathtt{y}}
\newcommand{\vz}{\mathtt{z}}
\newcommand{\vs}{\mathtt{c}}
\newcommand{\seq}{\;{=}\;}
\newcommand{\sett}{\;{:=}\;}
\newcommand{\Eqs}{\mathcal{E}\hspace{-2pt}\mathit{qs}}
\newcommand{\Neqs}{\mathcal{N}\hspace{-2.5pt}\mathit{eg}}
\newcommand{\SimRel}{\propto}
\newcommand{\RedRel}{\precapprox}
\newcommand{\TrRel}{\leadsto}
\newcommand{\ParSet}{\mathcal{P}}
\newcommand{\TreeData}{\mathscr{T}}
\newcommand{\GetParSet}{\mathsf{Pars}}
\newcommand{\VarSet}{\mathcal{V}}
\newcommand{\EParSet}{\mathcal{P}_{s}}
\newcommand{\SParSet}{\mathcal{P}_{c}}
\newcommand{\ifop}{\mathtt{if}}

\newcommand{\unsat}{\mathrm{unsat}}

\newcommand{\SetExpr}{\mathcal{E}}
\newcommand{\Predic}{\mathcal{P}\hspace{-2pt}\mathit{red}}
\newcommand{\Narr}{\nu}
\newcommand{\ConfSet}{\mathcal{C}}
\newcommand{\empt}{\varepsilon}

\newcommand{\OutTrue}{\mathbf{T}}
\newcommand{\OutFalse}{\mathbf{F}}
\newcommand{\rar}{\mapsto}
\newcommand{\funrar}{\rightarrow}
\newcommand{\mindia}{\rule{0mm}{1.63ex}}    
\newcommand{\ov}[1]{*+[F-:<44pt>]{#1\mindia}}
\newcommand{\greq}{\;{\equiv}\;}

\newtheorem{Definition}{Definition}

\newtheorem{Example}{Example}
\def\mscpa{\texttt{MSCP-A\,\,}}
\def\tlet{\textrm{let}\;}
\def\tin{\;\textrm{in}\;}
\def\logor{\mathrel{\vee}}
\def\logimpl{\mathrel{\Rightarrow}}
\def\logand{\mathrel{\&}}
\def\lognot{\mathop{\neg}}

\def\longconc{\scriptstyle{\mathrel{++}}\displaystyle}
\def\GenOp{\mathrm{Gen}}
\def\iff{\mathrel{\Leftrightarrow}}

\def\frameit{\smallskip \advance \linewidth by -7.5pt \setbox0=\vbox \bgroup
\strut \ignorespaces }

\begin{document}
\maketitle

\begin{abstract}
We describe a general parameterized scheme of program and constraint analyses allowing us to specify both the program specialization method known as Turchin's supercompilation and Hmelevskii's algorithm solving the quadratic word equations. The scheme is specified for both sorts of the analysis and works in a joint algorithm in which these two sorts of the analysis are used together. The word equations and the inequalities on regular patterns are used as the string constraint language in the algorithm. 

\end{abstract}

\section{Introduction}

Program transformation techniques are usually used for optimization, but sometimes they are also used for verification. Given a program $\Prog$, a simple analysis of its transformed version might show some properties of $\Prog$ which were not obvious in $\Prog$ itself \cite{Hamilton15,NemReachability}. This paper aims at verification of reachability properties of functional programs. Any program $\Prog$ can be formally unfolded in a tree which includes all computation paths of $\Prog$ \cite{Jones}. Some unreachable paths in the tree are pruned by the transformation. The resulted tree is presented by the unfold/fold transformation by a finite graph. If the graph does not contain some program state, then the state is unreachable. E.\,g., given $\Prog$ returning either $\OutTrue$ or $\OutFalse$, let the graph after the unfold/fold transformation \cite{Burst,Jones} contain no states with value $\OutFalse$. Then the value $\OutFalse$ is unreachable from the input point of $\Prog$: either the value $\OutTrue$ will be returned or $\Prog$ will run forever.

Turchin's supercompilation is one of such methods based on unfold/fold operations \cite{Tur86,Jones,Hamilton15}. Turchin's original works use the string operating language Refal as the input language of a supercompiler \cite{TurRefal5}. When a string operating program is treated by supercompilation, the method requires analysis of word equations.

\begin{Example}

Let a program include an $\ifop$ operator testing the equality $\vx\seq\vy$ on strings. Let $\pu$ take strings as its value, $\cA$ and $\cB$ be letters, and $\longconc$ --- the concatenation sign. Let $\vx\sett \pu\longconc\cA$, $\vy\sett \cB\longconc\pu$. In order to prove that the test $\vx\seq\vy$ never returns $\OutTrue$, we need to know that no value of $\pu$ can satisfy the equality $\pu\longconc\cA\seq\cB\longconc\pu$. That means the word equation $\pu\longconc\cA\seq\cB\longconc\pu$ has no solutions, or, in other words, the relation on strings defined by the equation $\pu\longconc\cA\seq\cB\longconc\pu$ is empty. 

\end{Example}

As a rule, tools for string programs analysis are based on the class of regular languages as a language for string constraints \cite{Yu,Trinh,Bjorner}. This class is then enriched by some additional predicate symbols in such a way that decidability of the enriched systems is preserved. The class of the word relations defined by the word equations is neither a subset nor a superset of the rational relation set of rational relations defined by the finite-state machines \cite{Plandowski}. 

This work considers the set of quadratic word equations as a constraint language for supercompilation. Given a program $\Prog$,  the word equations are used either to detect some of unreachable computation paths of $\Prog$ or to present some properties of the program structures.

Our contributions are the following:

\begin{enumerate}
\item We present a general parameterized scheme of program and constraint analyses allowing us to specify both a program specialization method known as Turchin's supercompilation and Hmelevskii's algorithm solving the quadratic word equations.
\item We specify this scheme for each of the two analyses and present a new joint algorithm in which these two analyses are used together. The joint algorithm verifies some safety properties of the programs to be analysed. A new type of string constraints, namely the quadratic word equations, together with the inequalities on regular patterns are used for the constraint analysis. The input string lengths of the programs are unknown, \ie they are not bounded in advance.
\item The presented algorithm has been implemented in a model supercompiler \mscpa for the language Refal. 
\end{enumerate}

The paper is organized as follows. After introducing the syntax (\sectionname~\ref{Subsection:Language}), we describe the general formal scheme of the analysis (\sectionname~\ref{Subsection:GeneralScheme}). Then we describe the set of configurations (\sectionname~\ref{subsection:constraints}) and specify the scheme first for solving quadratic word equations (\sectionname~\ref{section:ContstraintScheme}) and then for supercompilation (\sectionname~\ref{section:supercompilation}). Section~\ref{section:TwoSchemes} informally describes how these schemes are used together in the supercompiler \mscpa \hspace{-4pt}. We show some examples of reachability analysis done using the schemes in Appendix (\sectionname~\ref{section:examples}). 

\section{Presentation Language}\label{Subsection:Language}

We present our program examples in a variant of a pseudocode for functional programs. The  programs  given  below  are  written
as term rewriting systems based on pattern matching.  The rules in the programs
are  ordered  from  the top  to  the  bottom  to  be  matched. The programs are over strings in a finite alphabet $\Sigma$ and use variables of two types: strings, which can be valued by elements of $\Sigma^*$, and symbols taking elements of $\Sigma$. We use the notion of a parameter for data which already has a value but it is unknown to us; while a variable value is undefined and is to be assigned. We use the following syntax.

\begin{itemize}
\item $\empt$ is the empty word, $\longconc$ is the concatenation sign (both may be omitted);
\item $\cA$, $\cB$, $\cC$ \etc are elements of $\Sigma$; 
\item $\pu$, $\pv$, $\pw$, maybe subscripted, are string parameters; $\pt$ is a symbol parameter;
\item $\vx$, $\vy$, $\vz$, maybe subscripted, are string variables; $\vs_i$ is a symbol variable;
\item function names are given in typographic style and start with capital letters, \eg, $\BothA$, $\Iter$, $\FPal$.
\end{itemize}

Let $\VarSet$ denote a set of all variables, $\EParSet$ and $\SParSet$ a set of string and symbol parameters respectively, and $\ParSet\seq\EParSet\cup\SParSet$.

A rule of a program is $\Rule_l\seq\Rule_r$, where $\Rule_l$ is $\Fun(\Phi_1,\dots,\Phi_n)$, $\Fun$ is an $n$-ary function name, $\Phi_i\in (\Sigma\cup\VarSet)^*$, $\Rule_r$ can contain symbols, function calls, and variables. $\Rule_r$ may include only the variables present in $\Phi_1\longconc\dots\longconc\Phi_n$.

\begin{Definition}
Expression set $\SetExpr$ and function $\GetParSet(\Phi)\colon \SetExpr \funrar 2^{\ParSet}$ returning the set of parameters of an expression are defined as follows.
\begin{itemize}
\item If $\Phi\in\Sigma\cup\{\empt\}$, then $\Phi\in\SetExpr$. $\GetParSet(\Phi)\seq\emptyset$. If $\Phi\in\ParSet$, then $\Phi\in\SetExpr$, $\GetParSet(\Phi)\seq \{\Phi\}$.
\item If $\Phi_1\in\SetExpr$, $\Phi_2\in\SetExpr$, then $\Phi_1\longconc \Phi_2\in\SetExpr$ and $\GetParSet(\Phi_1\longconc \Phi_2)\seq \ParSet(\Phi_1)\cup\ParSet(\Phi_2)$.
\item If $\Phi_1,\dots,\Phi_n\in\SetExpr$, $f$ --- a function name of the arity $n$, then $f(\Phi_1,\dots,\Phi_n)\in \SetExpr$, $\GetParSet(f(\Phi_1,\dots,\Phi_n))\seq\bigcup^n_{i=1}\ParSet(\Phi_i)$.
\item $\SetExpr$ does not include any other elements. 
\end{itemize}
\end{Definition}
Given a program, the function $\Go$ serves as its input point.

Function $\xi\colon\SetExpr\funrar\SetExpr$ is called a substitution, if $\xi$ is a morphism on $\SetExpr$ preserving constants. Thus, any substitution is completely defined by its values on $\ParSet$. We write an application of substitution $\xi$ to $\Phi$ as $\Phi\xi$. We also assume that substitutions respect types, \ie for every $\pt\in\SParSet$ $\pt\xi\in \Sigma\cup\SParSet$. If for $1\leq i \leq n$ $\pu_i\xi\seq\Phi_i$ and $\forall q\in\ParSet\setminus \{\pu_1,\dots,\pu_n\}$ $q\xi\seq q$, then we write $\xi$ as $\pu_1\rar\Phi_1,\dots,\pu_n\rar\Phi_n$. Substitutions for $\VarSet$ are defined similarly.

In this paper the notion of a substitution is also extended to predicates as follows. Given an $n$-ary predicate $P$ and substitution $\xi$, $P\xi$ means a restriction of $P$ to the image of $\xi$, \ie $P\xi\seq P'$, \st $\forall \pu_1,\dots,\pu_m (P'(\pu_1,\dots\pu_m)\iff \exists \pv_1,\dots,\pv_n P(\pv_1\xi,\dots,\pv_n\xi))$.    

\section{Unfold/Fold Program Transformation Method}

This section presents a variant of the unfold/fold technique used by supercompilation \cite{Tur86,Jones} and is refined to the string data type with a class of word equations and inequalities used as a constraint language. First, we extend the unfold/fold scheme given in \cite{NemReachability} for a wider set of configurations. The scheme described is applicable both to the program data and word equations. Then we specify the relations controlling the unfold/fold process for the both types of data. 

\subsection{General Unfold/Fold Scheme}\label{Subsection:GeneralScheme}

Given a set of predicates $\mathcal{S}\seq\{P_i\}$, $\mathcal{S}\xi$ is a set of predicates equivalent to $\bigwedge_i P_i\xi$. The relation $\greq$ is the textual coincidence, $\logimpl$ and $\iff$ are logical connectives with the usual meaning. Given a tree $\TreeData$ and the edge $\Node\funrar\Node'$ in $\TreeData$, we say $\Node'$ is a child of $\Node$. A node $\Node$ is an ancestor of $\Node'$ (and $\Node'$ is a successor of $\Node$) if there exists a sequence of edges such that $\Node\funrar\Node_1\funrar\dots\funrar\Node'$. 

\begin{Definition}
\emph{A configuration} is a tuple $\langle \Phi, \Predic \rangle$, where $\Phi\in\SetExpr$, $\Predic$ is a set of predicates on $\GetParSet(\Phi)$. We denote the set of configurations by $\ConfSet$. 

Given $C\in\ConfSet$, and a substitution $\xi$, $C\xi\seq\langle \Phi\xi, (\Predic)\xi\rangle$.
\end{Definition}

\noindent Given a tuple $\TreeData\seq\langle C_0, \TrRel, \RedRel, \SimRel\rangle$, where $\TrRel$, $\RedRel$, and $\SimRel$ are binary relations on $\ConfSet$, and $C_0\in\ConfSet$, we name $\TrRel$ the transition relation, $\RedRel$ --- the reducing relation, $\SimRel$ --- the similarity relation. We assume that every node $\Node$ has a unique mark: either \textit{fresh}, \textit{open} or \textit{closed with $\Node'$}, where $\Node'$ is either an ancestor of $\Node$ or $\Node$ itself. The unfolding of the process tree of $\TreeData$ follows the scheme below, which is an extension of the scheme described in \cite{NemReachability}. 

\textbf{START:} Create a root $\Node_0$ of the tree, label it with $C_0$ (denoted $C(\Node_0)\seq C_0$) and mark is as fresh. 

\textbf{UNFOLD:} Choose a fresh vertex $\Node$ and
generate configurations $C_i$ such that $C(\Node) \TrRel C_i$ and for every substitution $\sigma\colon\ParSet\funrar\Sigma^*$, if $C\sigma \TrRel C_{\sigma}$, then there exist a substitution $\sigma'$ and $i$ such that $C_i\sigma'\greq C_{\sigma}$. For every such a $C_i$ create a fresh child
vertex $\Node_i$ and label it with $C_i$. Open the vertex $\Node$. If the parent $\Node'$ of $\Node$ is
open, then mark $\Node'$ by \textit{closed with $\Node'$}. 

\textbf{CLOSE I}: Choose an open vertex $\Node$ and check whether it has an
ancestor vertex $\Node'$, such that $C(\Node)\RedRel C(\Node')$. If yes mark the vertex $\Node$ as \textit{closed with $\Node'$} and delete the childen of $\Node$.

\textbf{CLOSE II}: Choose an open vertex $\Node$ and check whether all its children
are closed. If yes mark the vertex $\Node$ by \textit{closed with $\Node$}. If $C(\Node)\seq\langle\Phi,\Predic\rangle$ s.t. $\Phi\in (\Sigma\cup\ParSet)^*$, mark $\Node$ by \textit{closed with $\Node$}.

\textbf{GENERALIZE}: Choose an open vertex $\Node$ and its ancestor vertex $\Node'$ such that $C(\Node')\SimRel C(\Node)$. Let $C(\Node')\seq\langle \Phi,\Predic\rangle$. Generate configuration $C_g$ such that both $C(\Node)\RedRel C_g$ and $C(\Node')\RedRel C_g$ 
hold, and there is a substitution $\xi\colon q_i\rar\Phi_i$ such that $C_g \xi\seq C(\Node')$. Delete the subtree with the root $\Node'$, except the vertex $\Node'$ itself. Replace the label $C(\Node')$ with a special \textit{let}-label $\Lambda\seq$''$\tlet \xi \tin C_g$'' and generate $i+1$ fresh children nodes\footnote{The construction of the \textit{let}-branching differs from the branching produced by \textbf{UNFOLD}. The children of a \textit{let}-node are generated by the split procedure rather than transitions: they all must be computed for computing $C(\Node')$.} of $\Node'$. The last is labeled with $C_g$ and the others with $\langle \Phi_i, \Predic'_i\rangle$, where $\Predic'_i\subseteq\Predic$, $\GetParSet(\Predic'_i)\subseteq\GetParSet(\Phi_i)$. Nodes with the \textit{let}-label are never tested by \textbf{CLOSE I} and \textbf{GENERALIZE}.

\textbf{PRUNE}: Choose a closed vertex $\Node$ and consider the subtree $\TreeData_{\Node}$ rooted in $\Node$. If all leaves in $\TreeData_{\Node}$ are closed with their ancestors from $\TreeData_{\Node}$, then delete $\TreeData_{\Node}$, node $\Node$ itself and the ingoing edge to $\Node$.

\subsection{Equations and Inequalities as String Constraints}\label{subsection:constraints}

In this subsection we specify a set $\Predic$ which is used in configuration set $\ConfSet$, both for the program and constraint analyses. The set consists of the two subsets, namely word equations and word inequalities. 

\begin{Definition}
\emph{A word equation} is an equality $\Phi_1\seq\Phi_2$, where $\Phi_1,\Phi_2\in (\ParSet\cup\Sigma)^*$. $\GetParSet(\Phi_1\seq\Phi_2)$ is $\GetParSet(\Phi_1\longconc\Phi_2)$. 
 
Given an equation $E\colon\Phi_1\seq\Phi_2$, \emph{a solution} of $E$ is $\xi\colon\ParSet\funrar\Sigma^*$ s.\,t. $\Phi_1\xi \greq \Phi_2\xi$. $E$ is \emph{quadratic}, if no string parameter occurs in $E$ more than twice \cite{DiekertJewels}. 
\end{Definition}

\begin{Definition}
Given $\Phi_i\in (\Sigma\cup\SParSet)^*$, and $q\in\ParSet$, \emph{a linear word inequality} is an inequality $\Ineq$ of the form $\forall \vz_1,\dots, \vz_n (q\neq \Phi_0\conc \vz_1\conc \Phi_1\conc\dots\conc \vz_n\conc \Phi_n)$, where for $1\leq i\leq n-1$ $\Phi_i\neq\empt$, $\GetParSet(\Ineq)$ is $(\bigcup_i\GetParSet(\Phi_i))\cup\{q\}$. Recall that $\vz_i$ are variables of the string type (\sectionname~\ref{Subsection:Language}). 

For the sake of brevity, we use the simplified notation $q\neq\Phi_0\conc\vz_1\conc\dots\conc\vz_n\conc\Phi_n$ treating all $\vz_i$ as free variables.
\end{Definition}

In the model used in this paper, any configuration is of the form $\langle \Phi,\Predic\rangle$, where $\Predic\seq\Eqs\cup\Neqs$, $\Eqs$ is a set of the quadratic word equations, and set $\Neqs$ is a set of the linear inequalities.

\section{Scheme for Constraint Analysis}\label{section:ContstraintScheme}

In this section we apply the scheme given in \sectionname~\ref{Subsection:GeneralScheme} to analysis of the word equations. As a result, we reconstruct the well-known algorithm of Hmelevskii for solving the quadratic word equations \cite{Hmelevsky} in the terms of the scheme given in \sectionname~\ref{Subsection:GeneralScheme}. The algorithm is extended to parameters from $\SParSet$ and constraints in the form of the linear inequalities. In order to get the algorithm we have to specify some versions of the relations $\TrRel$ and $\RedRel$ and the configuration set $\ConfSet_{Eq}$. 

Given a binary constructor $\ConstrEq$, \emph{an eq-configuration} $C\in\ConfSet_{Eq}$ is the configuration $C\langle \ConstrEq(\Phi_1,\Phi_2),\Eqs,\Neqs\rangle$, where $\Phi_1,\Phi_2\in(\Sigma\cup\ParSet)^*$ and the set $\Eqs\cup\{\Phi_1\seq\Phi_2\}$ includes the only quadratic word equation (and maybe some more linear word equations). 

Now we specify the $\TrRel$ and $\RedRel$ relations over $\ConfSet_{Eq}$. Consider eq-configurations $C_1\seq \langle \ConstrEq(\Phi^1_1,\Phi^1_2), \Eqs_1, \Neqs_1\rangle$, $C_2\seq \langle \ConstrEq(\Phi^2_1,\Phi^2_2), \Eqs_2, \Neqs_2\rangle$.

\begin{Definition}

$C_2$ \emph{is reduced} to $C_1$ (denoted $C_2\RedRel_{eq} C_1$), if there is a substitution $\xi$ such that $\Phi^1_1\xi\greq \Phi^2_1$, $\Phi^1_2\xi\greq \Phi^2_2$, $\Eqs_2\logimpl(\Eqs_1)\xi$, $\Neqs_2\logimpl(\Neqs_1)\xi$, and if $\pu\in\EParSet$ then $\pu\xi\in\EParSet$, if $\pt\in\SParSet$ then $\pt\xi\in\SParSet$. Thus, $\xi$ is a renaming substitution.  

\end{Definition}
 
\begin{Definition}
\label{Definition:EquationTransition}

$C_2$ is unfolded from $C_1$ (denoted $C_1\TrRel_{eq} C_2$), if there is a substitution $\Narr$ (which is called the narrowing substitution) satisfying $(\Eqs_1)\Narr\logimpl \Eqs_2$, $(\Neqs_1)\Narr\logimpl\Neqs_2$, and having the following properties.

Let $\Phi^1_1\seq q\conc\Phi'$, where $q\in\ParSet$, $q'\in\EParSet$, $\pt\in\SParSet$, $q'$ and $\pt$ are fresh parameters. If $\Phi_2^1\seq q\conc\Psi$, where $q\in\ParSet$, then all the definitions should be applied to $q$ symmetrically.
\begin{itemize}
\item If $\Phi_2^1\seq\empt$ and $q\in\EParSet$, then $\Narr\colon q\rar\empt$.
\item If $\Phi_2^1\seq \pu$ and $q,\pu\in \EParSet$, then $\Narr\colon q\rar\pu$.
\item If $\Phi_2^1\seq \pu\conc\pv\conc\Psi$ and $q,\pu,\pv\in \EParSet$, then either $\Narr\colon q\rar\pu\conc\pt\conc q'$ or $\Narr\colon q\rar\pu$.
\item If $\Phi_2^1\seq\pu\conc r\conc\Psi$, $r\in \SParSet\cup\Sigma$, and $q\in \EParSet$, then either $\Narr\colon q\rar\pu\conc r\conc q'$ or $\Narr\colon q\rar\pu$.
\item If $\Phi_2^1\seq r\conc\Psi$, $r\in \SParSet\cup\Sigma$, and $q\in \EParSet$, then either $\Narr\colon q\rar r\conc q'$ or $\Narr\colon q\rar\empt$.
\item If $\Phi_2^1\seq r\conc\Psi$, $r\in \SParSet\cup\Sigma$ and $q\in \SParSet$, then $\Narr\colon q\rar r$.
\end{itemize}

Let $\Phi^1_1\Narr\seq\Psi_0\conc\Psi_1$, $\Phi^1_2\Narr\seq\Psi_0\conc\Psi_2$, and $\Psi_0$ is chosen to be of the maximal length. Then assign $\Phi^2_1\sett \Psi_1$, $\Phi^2_2\sett \Psi_2$. 
\end{Definition}

Actually, expression $\ConstrEq(\Phi^2_1,\Phi^2_2)$ presents the equation $\Phi^1_1\Narr\seq\Phi^1_2\Narr$ after deleting the common prefixes of $\Phi^1_1\Narr$ and $\Phi^2_1\Narr$. The construction of $\Neqs_2$ from $\Neqs_1\Narr$ is shown in the Appendix (\sectionname~\ref{subsection:negrules}). The properties of $\RedRel_{eq}$ and $\TrRel_{eq}$, together with the construction of $\Neqs_2$, guarantee that the algorithm given in \sectionname~\ref{Subsection:GeneralScheme} terminates. In fact, this algorithm specified by $\RedRel_{eq}$ and $\TrRel_{eq}$ is a version of Hmelevskii's algorithm solving the quadratic word equations \cite{Hmelevsky} with some minor changes due to extension of the parameters set by $\SParSet$. 

In the unfolded process tree of an equation, some simple properties holding for every path generated in the tree may become explicit. If the properties are expressible as narrowings of the root parameters, then the narrowings are extracted from the tree and are used in the analysis in the program from where constraints come. The unfolding also performs a test for satisfiability of the equation $\Phi_1\seq\Phi_2$ under the conditions $\Eqs\logand\Neqs$. If the tree has no leaves marked by the expression $\ConstrEq(\empt,\empt)$ (which is replaced by $\OutTrue$ in our diagrams) then the equation $\Phi_1\seq\Phi_2$ has no solutions under the given conditions and the node with the general configuration $C\seq\langle\Psi,\Predic\rangle$ where $(\Eqs\cup\{\Phi_1\seq\Phi_2\}\cup \Neqs)\subseteq\Predic$, can be pruned.

An example of the constraint analysis following the scheme above is shown in Appendix, see Example~\ref{Example:EquationSolving}.

\section{Scheme for Program Analysis}\label{section:supercompilation}

Now we specify versions of the relations $\TrRel$, $\RedRel$, $\SimRel$ used by our program analysis. Consider a program $\Prog$, which is a finite sequence of rules $\Rule_l\seq \Rule_r$ (see \sectionname~\ref{Subsection:Language}).

\begin{Definition}
The homeomorphic embedding $\trianglelefteq$ is defined on $\SetExpr$ as follows \cite{Jones}.

\begin{itemize}
\item For every $\pu_1, \pu_2 \in \EParSet$, $\pt_1, \pt_2\in \SParSet$, $\pu_1\trianglelefteq\pu_2$, $\pt_1\trianglelefteq \pt_2$. For every $\Phi\in\SetExpr$, $\Phi\trianglelefteq \Phi$.
\item Given $\Phi_i,\Psi_i\in \SetExpr$, if $\Phi_1\trianglelefteq \Phi_2$, then $\Phi_1\trianglelefteq \Psi_1\conc\Phi_2\conc\Psi_2$ (for any $\Psi_1, \Psi_2\in\SetExpr$), and $\Phi_1\trianglelefteq f(\Psi_1,\dots,\Psi_n)$, where for some $i$ $\Psi_i\seq\Phi_2$.
\item Given $\Phi_i,\Psi_i\in \SetExpr$, if $\forall i(\Phi_i\trianglelefteq \Psi_i)$ then $f(\Phi_1,\dots,\Phi_n)\trianglelefteq f(\Psi_1,\dots,\Psi_n)$.
\end{itemize}

\end{Definition}

Let two configurations $C_1,C_2\in\ConfSet$ be $C_i\seq\langle \Phi_i,\Eqs_i,\Neqs_i\rangle$.

\begin{Definition}

$C_2$ \emph{is reduced} to configuration $C_1$ (denoted $C_2\RedRel C_1$), if there is a substitution $\xi$ such that $\Phi_1\xi\greq \Phi_2$, $\Eqs_2\logimpl(\Eqs_1)\xi$, $\Neqs_2\logimpl(\Neqs_1)\xi$. 

\noindent $C_2$ \emph{is similar} to $C_1$ (denoted $C_1\SimRel C_2$) if $\Phi_1\trianglelefteq\Phi_2$ .
\end{Definition}

\begin{Definition}
$C_1$ \emph{is unfolded} to $C_2$ ($C_1\TrRel C_2$) if there exists a rule $\Rule_l\seq\Rule_r$ in $\Prog$ such that there are a substitution $\Narr\colon \pu_i\rar\Psi_i$ and a set of equations $\Eqs_{Narr}$ over $\bigcup_i\GetParSet(\Psi_i)$ such that $\exists \sigma\colon\VarSet\funrar\SetExpr((\Eqs_{Narr}\iff \Rule_l\sigma\greq \Phi_1\Narr)\logand \Rule_r\sigma\greq\Phi_2)$. We call $\Narr$ the narrowing, and the elements of $\Eqs_{Narr}$ the narrowing equations. Moreover, the following properties are required.
\begin{itemize}
\item $\Neqs_2\seq \Neqs_{Subst}\cup \Neqs_{Scr}$ such that $(\Neqs_1)\Narr\logimpl \Neqs_{Subst}$, and, for every inequality $\Ineq\in\Neqs_{Scr}$, if $\lognot\Ineq$, then there is a rule $\Rule'_l\seq\Rule'_r$ such that it precedes $\Rule_l\seq\Rule_r$ and $\forall \sigma\colon\VarSet\funrar\SetExpr.\,(\Rule_l\sigma\greq(\Phi_1)\Narr\logimpl \exists \sigma'\colon\VarSet\funrar\SetExpr.\,(\Rule'_l\sigma\greq(\Phi_1)\Narr))$. 
\item $\Eqs_2\seq\Eqs'_{Narr}\cup\Eqs_{Subst}$ such that $(\Eqs_1)\Narr\logimpl\Eqs_{Subst}$, $\Eqs_{Narr}\logimpl\Eqs'_{Narr}$, and all equations in $\Eqs_2$ are quadratic.  
\end{itemize} 
\end{Definition}

Actually, $\Eqs_{Narr}$ makes sense only if  $\Rule_l$ contains multiple occurrences of some string variables. Because the rules of $\Prog$ are ordered from top to bottom to be matched, the branches of the process tree generated by $\textbf{UNFOLD}$ rule are ordered. The set $\Neqs_{Scr}$ is constructed using this order. The order is not used in our analysis except this case.  
 
Unlike the scheme given in \cite{NemReachability}, the unfolding scheme in this paper only partially determines the transitions done by \textbf{UNFOLD}, for they may vary in the equation and inequality sets. A construction demonstrating the role of $\Eqs_{Narr}$ and the problem to make $\Neqs_{Scr}$ explicit is given in Appendix (Example~\ref{Example:NarrEqNarrNeq}).  

\begin{Example}\label{Example:FPalDef}

The function $\FPal$ below tests whether its argument is a palindrome.

$$
\xymatrix@-4.5mm {
*+[F-]{\begin{array}{llll}
\Rule^{\FPal}_1: & \FPal(\vs\conc \vx\conc \vs)&=&\FPal(\vx);\\
\Rule^{\FPal}_2: & \FPal(\vs_1\conc\vx\conc\vs_2)&=&\OutFalse;\\
\Rule^{\FPal}_3: & \FPal(\vs)&=&\OutTrue;\\
\Rule^{\FPal}_4: & \FPal(\empt)&=&\OutTrue;\\ 
\end{array}
}}
$$

Given configuration $C\seq\langle\FPal(\pu),\cA\conc\pu\seq\pu\conc\cA,\OutTrue\rangle$ and the configurations 

$$
\begin{array}{ll}
C^{I} & \seq\langle\FPal(\pu_1),\cA\conc\pu_1\seq\pu_1\conc\cA,\OutTrue\rangle \\
C^{II}& \seq\langle \FPal(\pu_1),\cA\conc\pu_1\seq\pw\conc\cA,\OutTrue\rangle\\
C^{III}&\seq\langle\FPal(\pu_1),\OutTrue,\OutTrue\rangle
\end{array}
$$

\noindent all the relations $C\TrRel C^{I}$, $C\TrRel C^{II}$, and $C\TrRel C^{III}$ hold. The corresponding rule is $\Rule^{\FPal}_1$, the narrowing substitution is $\Narr\colon\pu\rar\cA\conc\pu_1\conc\cA$.  
\end{Example}

\begin{Definition}
\emph{A generalization of $\Phi_1,\Phi_2\in\SetExpr$} is an expression $\GenOp(\Phi_1, \Phi_2)\seq\Psi$ such that there are $\xi_1, \xi_2\colon \ParSet\funrar\SetExpr$, named \emph{generalizing substitutions}, such that $\Psi\xi_i\greq \Phi_i$ ($i\seq 1, 2$).

\emph{A generalization of a linear inequality} $q \neq \Psi(\vz_1,\dots,\vz_n)$ is either $\OutTrue$ or an inequality $q \neq \Psi(\vz_1,\dots,\vz_n)\sigma$, where $\sigma$ maps some of $\vz_i$ to constant strings (maybe empty) and preserves the others.

\emph{A generalization of a word equation} $\Phi_1(\pu_1,\dots,\pu_n)\seq\Phi_2(\pu_1,\dots,\pu_n)$ is either $\OutTrue$ or a quadratic word equation $\Phi'_1(\pu_1,\dots,\pu_{n+k})\seq \Phi'_2(\pu_1,\dots,\pu_{n+k})$ such that there is $\xi\colon \{\pu_{n+1},\dots,\pu_{n+k}\}\funrar\{\pu_1,\dots,\pu_n\}\cup{\Sigma}^+$ satisfying $\Phi'_i \xi\greq \Phi_i$ ($i\seq 1, 2$).
\end{Definition}

\begin{Example}
Given inequality $\Ineq\colon \pu\neq \vz_1\conc\cA\conc\vz_2$, all the inequalities $\pu\neq\vz\conc\cA$, $\pu\neq\vz\conc\cA\conc\cB$, $\pu\neq\cA$ are generalizations of $\Ineq$. Inequality $\pu\neq\vz'_1\conc\cA\conc\vz'_2\conc\cB$ is not a generalization of $\Ineq$, because substitution $\vz_2\rar\vz'_2\conc\cB$ is non-constant.

Given equation $E\colon \cA\conc\pu\seq\pu\conc\cA$, all the equations $\cA\conc\pu\seq\pw\conc\cA$, $\cA\conc\pw\seq\pu\conc\cA$, and $\pv\conc\pu\seq\pu\conc\pv$ are generalizations of $E$. The equation $\pw\conc\cA\conc\pu\seq\pu\conc\cA\conc\pw$ is not a generalization of $E$, because substitution $\pw\rar\empt$ is forbidden.  
\end{Example}

\begin{Definition}
\emph{A generalization of two configurations} $C_1\seq \langle \Phi_1, \Eqs_1, \Neqs_1\rangle$, $C_2\seq\langle \Phi_2,\Eqs_2,\Neqs_2\rangle$ is $C_g\seq \langle \Phi_g,\Eqs_g,\Neqs_g\rangle$ such that

\begin{itemize}
\item $\Phi_g\seq \GenOp(\Phi_1,\Phi_2)$, $\xi_1$ and $\xi_2$ are the generalizing substitutions, $\Phi_g\xi_i\greq\Phi_i$.
\item $\Neqs_1\logimpl(\Neqs_g)\xi_1$, $\Neqs_2\logimpl (\Neqs_g)\xi_2$ and for all $\Ineq(\pu)\in\Neqs_g$ such that $\pu\xi_1\notin \Sigma^*$, there exists $\Ineq'\in\Neqs_1$ such that $\Ineq(\pu)$ is a generalization of $\Ineq'\xi_1$.
\item $\Eqs_1\logimpl (\Eqs_g)\xi_1$, $\Eqs_2\logimpl(\Eqs_g)\xi_2$ and for all $E \in\Eqs_g$ such that $\exists \pu(\pu\in\GetParSet(E)\logand \pu \xi_1 \notin \Sigma^*)$, there exists an equation $E'\in\Eqs_1$ such that $E$ is a generalization of $E'\xi_1$. 
\end{itemize}
\end{Definition}

Thus the set of the computation paths starting at the generalized configuration $C_g$ includes all computation paths both from $C_1$ and $C_2$.

\begin{Example}
\label{Example:FPalSolve}

Consider the function $\FGenA$ below generating a word in language $\cA^*$. Function $\FPal$ is defined in Example~\ref{Example:FPalDef}. 

Given the input point $\Go(\pw)\seq \FPal(\FGenA(\pw))$, let us prove that the configuration $\langle\OutFalse,\Eqs,\Neqs\rangle$ is unreachable using the unfold/fold scheme above. If $\Neqs$ or $\Eqs$ consists of a single element, then we omit the set enclosing brackets.

$$
\xymatrix@-4.5mm {
*+[F-]{\begin{array}{llll}
\Rule^{\FGenA}_1: & \FGenA(\empt)&=&\empt; \\
\Rule^{\FGenA}_2: & \FGenA(\vs\conc\vx)&=&\cA\conc \FGenA(\vx);
\end{array}}}
$$

The input configuration is $C_0\seq \langle \FPal(\FGenA(\pw)),\OutTrue,\OutTrue\rangle$. \textbf{\emph{UNFOLD}} rule is applied to the input node. The first child corresponding to the application of $\Rule^{\FGenA}_1$ is generated by narrowing $\pw\rar\empt$ and is labeled with $C_1\seq \langle \FPal(\empt), \OutTrue, \OutTrue\rangle$. The second child corresponding to the application of $\Rule^{\FGenA}_2$ is generated by narrowing $\pw\rar\pt\conc\pw_1$ and is labeled with $C_2\seq \langle \FPal(\cA\conc \FGenA(\pw_1)),\OutTrue,\OutTrue\rangle$. After applying \textbf{UNFOLD} to $C_2$, no contradiction is generated. The similarity relation $C_0\SimRel C_2$ holds, therefore \textbf{\emph{GENERALIZE}} is used. 

The first steps (except unfolding $C_2$) are shown in the diagram below. For the sake of brevity, we omit the trivial equations and inequalities sets in the configurations.
$$
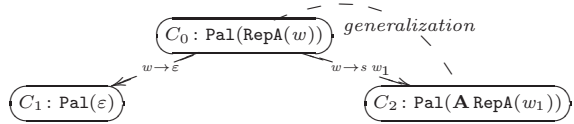
\scriptsize
\xymatrix @-4mm {
&\ov{C_0\colon \FPal(\FGenA(\pw))}\ar[dl]|{\skipper\pw\rightarrow\empt}\ar[dr]|{\skipper\pw\rightarrow\pt\conc\pw_1}\\
\ov{C_1\colon \FPal(\empt)}&&\ov{C_2\colon \FPal(\cA \conc\FGenA(\pw_1))}\ar@{--}@/_2pc/|<(.3){\txt{generalization}}[lu]
}
$$

The following generalization with generalizing substitutions $\xi_1\colon \{\pu_1\rar\empt, \pu_2\rar\pw\}$, $\xi_2\colon \{\pu_1\rar\cA, \pu_2\rar\pw_1\}$ is built.

$\tlet \pu_1=\empt,\,\pu_2=\pw \tin C_g\seq\{\FPal(\pu_1\conc\FGenA(\pu_2)), \pu_1\conc\cA\seq\cA\conc\pu_1, \OutTrue\}$.

The equation $E\colon \pu_1\conc\cA\seq\cA\conc\pu_1$ appears in $C_g$. Both $\OutTrue\logimpl E\xi_1$ and $\OutTrue\logimpl E\xi_2$ hold, and $\pu_1\xi_1\in\Sigma^*$, therefore $E$ is a valid equation for $C_g$ (the rule generating the equation is given in Appendix, \sectionname~\ref{subsection:genrules}).

Then again, node $C_g$ generates two successors by applying \textbf{\emph{UNFOLD}}. First, consider $C^2_g\seq\langle\FPal(\pu_1\conc \cA\conc\FGenA(\pu_3)), \pu_1\conc\cA\seq\cA\conc\pu_1,\OutTrue\rangle$ corresponding to the narrowing $\pu_2\rar\pt\conc\pu_3$. Given substitution $\sigma\colon \pu_1\rar\pu_1\conc\cA,\,\pu_2\rar\pu_3$, it satisfies both $\FPal(\pu_1\conc \cA\conc\FGenA(\pu_3))\seq\FPal(\pu_1\conc \FGenA(\pu_2))\sigma$ and $(\pu_1\conc\cA\seq\cA\conc\pu_1)\sigma \logimpl \pu_1\conc\cA\seq\cA\conc\pu_1$. Hence, $C^2_g\RedRel C_g$ and \textbf{CLOSE I} marks $C^2_g$ as \textit{closed with} $C_g$.

Rule $\Rule^{\FPal}_1$ unfolds the configuration $C^1_g$ to $C^3_g\seq\langle\FPal(\pu_2),\pu_2\conc\cA\seq\cA\conc\pu_2,\OutTrue\rangle$ by means of the narrowing $\pu_1\rar \cA\conc\pu_2\conc\cA$. $C^3_g\RedRel C^1_g$ with the substitution $\sigma\colon \pu_1\rar\pu_2$ and \textbf{CLOSE I} marks $C^3_g$ as \textit{closed with} $C^1_g$.  

Rule $\Rule^{\FPal}_2$ unfolds $C^1_g$ to $C^4_g\seq\langle\OutFalse,\pt_1\conc\pu_3\conc\pt_2\conc\cA\seq\cA\conc\pt_1\conc\pu_3\conc\pt_2,\pt_1\neq\pt_2\rangle$ by means of $\pu_1\rar\pt_1\conc\pu_3\conc\pt_2$. The constraint analysis of this configuration shows the predicate $\pt_1\neq\pt_2\logand \pt_1\conc\pu_3\conc\pt_2\conc\cA\seq\cA\conc\pt_1\conc\pu_3\conc\pt_2$ is contradictory. Thus, the branch with the node labeled with $C^4_g$ is pruned. This transformation is crucial for solving the verification task given above.

Rules $\Rule^{\FPal}_3$ and $\Rule^{\FPal}_4$ unfold $C^1_g$ to the constant configurations $C^5_g$ and $C^6_g$, which are closed by \textbf{CLOSE II}. The whole process tree we constructed for the computation is shown below. If $\Node$ is closed with its ancestor $\Node'$, it has the dotted reverse edge to $\Node'$ marked ''\textit{reducing}''. If $\Node$ is closed with itself, we do not write the dotted edge corresponding to its mark.

$$
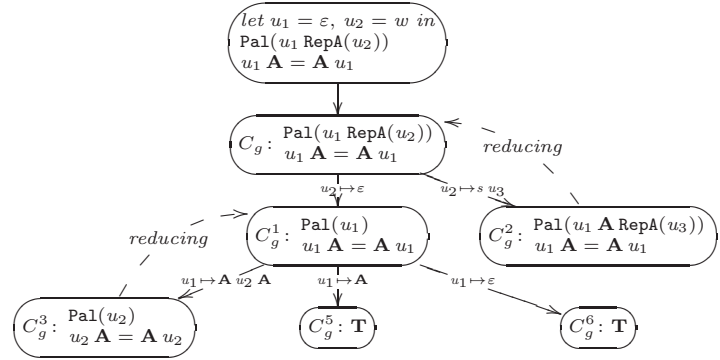
\scriptsize
\xymatrix @-4mm {
&\ov{\begin{array}{l}\tlet \pu_1=\empt,\,\pu_2=\pw \tin \\
\FPal(\pu_1\conc\FGenA(\pu_2)) \\
\pu_1\conc\cA\seq\cA\conc\pu_1
\end{array}}\ar[d]\\
&\ov{C_g\colon\begin{array}{l}\FPal(\pu_1\conc\FGenA(\pu_2))\\\pu_1\conc\cA\seq\cA\conc\pu_1
\end{array}}\ar[d]|{\skipper\pu_2\rar\empt}\ar[dr]|{\skipper\pu_2\rar\pt\conc \pu_3}\\
&\ov{C^1_g\colon \begin{array}{l}
\FPal(\pu_1) \\
\pu_1\conc\cA\seq\cA\conc\pu_1
\end{array}}\ar[dl]|{\skipper\pu_1\rar\cA\conc\pu_2\conc\cA}\ar[d]|{\skipper\pu_1\rar\cA}\ar[dr]|{\skipper\pu_1\rar\empt}
&\ov{C^2_g\colon\begin{array}{l}\FPal(\pu_1\conc \cA\conc\FGenA(\pu_3))\\ 
\pu_1\conc\cA\seq\cA\conc\pu_1\end{array}}\ar@{-->}@/_2pc/|<(.3){\txt{reducing}}[lu]\\
\ov{C^3_g\colon \begin{array}{l}
\FPal(\pu_2) \\
\pu_2\conc\cA\seq\cA\conc\pu_2
\end{array}}\ar@{-->}@/^2pc/|<(.3){\txt{reducing}}[ru]&\ov{C^5_g\colon \OutTrue}&\ov{C^6_g\colon\OutTrue}
}$$

\noindent The process tree does not contain nodes labeled with $\OutFalse$ expression. Hence, the output $\OutFalse$ is unreachable from the input point $\Go(\pw)\seq \FPal(\FGenA(\pw))$. The corresponding safety property of the program $\Prog$ has been proven.

\end{Example}

\section{Combining the Two Schemes}\label{section:TwoSchemes}

Let a configuration be $C\seq\langle \Phi,\Eqs,\Neqs\rangle$, the narrowing substitution be $\Narr$. To check whether $C\Narr$ is unreachable, we follow the algorithm below.

\begin{itemize}
\item Successively select elements of $(\Neqs)\Narr$ and replace them by their corollaries, which are linear inequalities (following the table given in Appendix, \sectionname~\ref{subsection:negrules}). If some $\Ineq\in(\Neqs)\Narr$ implies contradiction, then delete the node labeled with $C\Narr$. Otherwise construct the inequality set $\Neqs'$ for $C\Narr$.
\item Successively select elements of $(\Eqs)\Narr$ and split them into a number of shorter equations using the length argument \cite{Huova}. If the length argument for some $E\in(\Eqs)\Narr$ implies contradiction, delete the node labeled with $C\Narr$. Otherwise construct the set $\Eqs'$ of the simplified equations.
\item Successively select $E\in\Eqs'$. If $E$ is not quadratic, then generalize $E$ by $E'\colon\Phi_1\seq\Phi_2$. Take all elements $\Ineq_i$ of $\Neqs'$ such that $\GetParSet(\Ineq_i)\subseteq\GetParSet(\Phi_1\longconc\Phi_2)$, and all linear equations $E_j\in\Eqs'$ \st $\GetParSet(E_j)\cap\GetParSet(E')\neq\emptyset$. Apply our constraint analysis scheme (\sectionname~\ref{section:ContstraintScheme}) for $C_{eq}\seq\{\ConstrEq(\Phi_1,\Phi_2),\{E_j\},\{\Ineq_i\}\}$. Proceed until $\Eqs'$ is completely exhausted or the contradiction is found.

\end{itemize}

To check whether $\Neqs_1\logimpl\Neqs_2$, it is enough to check whether $\lognot\Neqs_2\logimpl\lognot\Neqs_1$. Because all elements $\Ineq_j\in\Neqs_2$, $\Ineq'_i\in\Neqs_1$ are linear and the cardinality of $\Sigma$ is more than 3, the implication always can be proved or refuted by finding whether for every $\lognot\Ineq_j$ there are such $i$, $\sigma\colon\VarSet\funrar(\VarSet\cup\Sigma)^*$ that $\lognot\Ineq_j \greq \lognot\Ineq'_i\sigma$ \cite{JainRegPatterns}.

The most complex problem in the analysis is checking whether $\Eqs_1\logimpl \Eqs_2$, even for sets of the quadratic equations. In general, the language inclusion problem for word equations is undecidable \cite{FreydenbergerBook}. To check the reducing relation for nodes labeled with $C_1\seq\langle\Phi_1,\Eqs_1,\Neqs_1\rangle$ and $C_2\seq\langle\Phi_2,\Eqs_2,\Neqs_2\rangle$, we must check whether $\Eqs_2\logimpl (\Eqs_1)\sigma$, and $(\Eqs_1)\sigma$ can contain non-quadratic word equations. In fact, our algorithm does simplify the equations of $(\Eqs_1)\sigma$ and then verifies  whether the simplified set is a subset of $Eqs_2$. Using such a simple test leads to more applications of \textbf{GENERALIZE} instead of \textbf{CLOSE I} when constructing the process tree, which results in a less precise analysis.

\section{Conclusion}

Unlike approaches shown in \cite{Bjorner,CVC4,Saxena}, our algorithm works for unbounded strings in the language of non-linear word equations. Attempts to replace the word equation languages by their regular approximations showed that all equation languages except languages described by very simple linear equations cannot be modelled by rational relations \cite{Yu}. Hence, using word equations as a constraint language can make sense. While in Example~\ref{Example:FPalSolve}, a regular condition $\pu_1\conc\cA\seq\cA\conc\pu_1$ is introduced, in Example~\ref{Example:irreg} and Example~\ref{Example:Sweden} verification is done on the non-regular data set.

The main two weaknesses of our approach are the following. First, the information about the branch ordering is lost, hence, if some subtree $\TreeData'$ of the process tree is cut by \textbf{PRUNE} rule, the computation paths which end in $\TreeData'$ may be embedded in some other subtree of $\TreeData$. Second, we weaken the constraints on parameters of the input \via generalization and reducing. 

\section*{Acknowledgements}
Without A.~P.~Nemytykh, who leaded the research and helped much to improve the paper, this paper would not exist.

\label{sect:bib}


\section*{Appendix}
\label{section:Appendix}

\subsection{Transition Rules}\label{subsection:negrules}

In the table below, if $\Phi\seq r\conc\Phi'$ then $\Phi[1]\seq r$, $\Phi[2]\seq \Phi'[1]$. The disjunction operation means that a branching in the process tree is generated. 

$$
\footnotesize
\begin{array}{l|l|l}
\text{Inequality}& \text{Narrowing Type} &\quad\text{Corollaries} \\ \hline
\pu\neq\vz_1\conc\Phi\conc\vz_m& \pu\rar\cA\conc\Psi & \Psi\neq\vz_1\conc\Phi\conc\vz_m\\
&&\Phi[1]\seq\cA \logimpl \Psi\neq\Phi[2]\dots\Phi[k]\conc\vz_m\\
&\pu\rar\pt\conc\Psi & \Psi\neq\vz_1\Phi\conc\vz_m\\
&&\pt\neq \Phi[1] \logor \Psi\neq \Phi[2]\dots\Phi[k]\conc\vz_m\\
&\pu\rar\pw\conc\Psi & \pw\neq \vz_1\conc\Phi\conc\vz_m\\
&&\Psi\neq\vz_1\conc\Phi\conc\vz_m\\
&\pu\rar\empt & \OutTrue\\
\pu\neq\Phi_0\vz_1\conc\Phi\conc\vz_n& 
\pu\rar\cA\conc\Psi& \Phi_0[1]\neq \cA\logimpl \OutTrue \\
&&\Phi_0[1]\seq \cA\logimpl \Psi\neq \Phi_0[2]\dots\vz_1\conc\Phi\conc\vz_n\\
&\pu\rar\pt\conc\Psi & \pt\neq\Phi_0[1]\logor \Psi\neq \Phi_0[2]\dots\vz_1\conc\Phi\conc\vz_n\\
&\pu\rar\pw\conc\Psi & \pw\neq \Phi_0\conc\vz_1\conc\Phi\conc\vz_m\\
&&\pw\neq\Phi_0\conc\vz\logor \Psi\neq \vz_1\conc\Phi\conc\vz_m\\
&\pu\rar\empt & \OutTrue\\
\pu\neq\vz_1\conc\Phi\conc\vz_n\conc\Phi_n& 
\pu\rar\cA\conc\Psi& \Psi\neq \vz_1\conc\Phi\conc\vz_n\conc\Phi_n \\
&&\Phi[1]\seq \cA\logimpl \Psi\neq \Phi[2]\dots\vz_n\conc\Phi_n\\
&\pu\rar\pt\conc\Psi & \Psi\neq \vz_1\conc\Phi\conc\vz_n\conc\Phi_n\\
&&\pt\neq\Phi[1]\logor \Psi\neq \Phi[2]\dots\vz_n\conc\Phi_n\\
&\pu\rar\pw\conc\Psi & \Psi\neq \vz_1\conc\Phi\conc\vz_n\conc\Phi_n\\
&&\pw\neq\vz_1\conc\Phi\conc\vz_n\logor \Psi\neq \vz\conc\Phi_n\\
&\pu\rar\empt & \OutTrue\\
\pu\neq\Phi_0\conc\vz_1\conc\Phi\conc\vz_n\conc\Phi_n& 
\pu\rar\cA\conc\Psi&\Phi_0[1]\neq \cA\logimpl\OutTrue\\
&&\Phi_0[1]\seq \cA\logimpl \Psi\neq \Phi_0[2]\dots\vz_1\conc\Phi\conc\vz_n\conc\Phi_n\\
&\pu\rar\pt\conc\Psi & \pt\neq\Phi_0[1]\logor \Psi\neq \Phi_0[2]\dots\vz_1\conc\Phi\conc\vz_n\conc\Phi_n\\
&\pu\rar\pw\conc\Psi & (\pw\neq\Phi_0\conc\vz\logand \Psi\neq \vz\conc\Phi_n) \logor \Psi\neq \vz_1\conc\Phi\conc\vz_n\conc\Phi_n\\
&&\qquad\logor\pw\neq\Phi_0\conc\vz_1\conc\Phi\conc\vz_n \\
&\pu\rar\empt & \OutTrue\\

\end{array}
$$

The implications above lose information, in particular, if the middle expression $\Phi$ contains some free variables. If $\Phi$ is long, splitting it into all possible parts produces too many branches in the process tree.

If $\Psi\seq q\longconc\Psi'$ and a corollary contains an inequality on $\Psi$, then the transition rules are applied successively until the first branching is generated. If an application of a transition rule produces the second branching, then the inequality on $\Psi$ producing the branching is thrown away. In the constraint analysis (\sectionname~\ref{section:ContstraintScheme}) such a situation never occurs, because of the restricted form of the narrowings.

Given an inequality $\pu\neq \Psi$, all its corollaries are of the form $\xi\neq\Psi'$, where $\Psi'$ is a substring of $\Psi$ (modulo renamings of $\vz_i$). 

\subsection{Special Generalizations}\label{subsection:genrules}

Here $\unsat(\Phi,\Psi)$ means the substitution of $\Phi$ in the inequality $\Psi$ results in the contradiction.

Rules for the inequalities are given below. For the third rule, besides the constraint $\pu\neq\Psi$, the special condition $\unsat(\Phi_2\conc\pu\conc\Phi_3,\Psi)$ must be checked.

$$
\footnotesize
\begin{array}{l|l|l|l}
\text{Ancestor}\skipper&\text{Successor}&\text{Conditions}&\text{Generalization}\\ \hline 
\empt & \pt & \pt\neq \cA & \pw\neq \vz_1\conc\cA\conc\vz_2\\
\empt & \pu & \pu\neq \Phi & \pw\neq\Phi \\
\Phi_1\in \Sigma^* & \Phi_2\conc\pu\conc\Phi_3 & \pu\neq \Psi \logand \unsat(\Phi_2\conc\pu\conc\Phi_3,\Psi)\logand \unsat(\Phi_1,\Psi) &\pw\neq\Psi
\\ \hline
\end{array}$$

Rules for the equations are given below. $()$ is a special delimiter symbol, and we assume $\pw,\pv_3,\pv_4,\Phi\neq\vz_1\conc ()\conc\vz_2$.
$$
\footnotesize
\begin{array}{l|l|l|l}
\text{Ancestor}\quad\skipper&\text{Successor}&\text{Conditions}&\text{Generalization}\\ \hline
\empt & \Phi\in(\Sigma\cup\ParSet)^+ & &\pw\conc\Phi\seq\Phi\conc\pw \\
\empt &\Phi\in(\Sigma\cup\ParSet)^+ & &\pv_1\conc ()\conc\pw\conc ()\conc\pv_2\seq ()\conc()\conc\pv_3\conc\Phi\conc\pv_4\conc()\\
\pu_1 & \Phi_1\conc\pu_2\conc\Phi_2 &\pu_i\conc\Psi\seq\Psi\conc\pu_i (i\seq 1,2) &\pw\conc\Psi\seq\Psi\conc\pw \\
 &  &\Phi_i\conc\Psi\seq\Psi\conc\Phi_i (i\seq 1,2)& 
\\ \hline
\end{array}
$$

The second rule for the equations in the table above is used optionally, for it generates new parameters $\pv_i$.

\subsection{Examples}\label{section:examples}

\begin{Example}\label{Example:EquationSolving}

For the sake of brevity, we write $\ConstrEq(\Phi_1,\Phi_2)$ as $\underline{\Phi_1\seq\Phi_2}$ and omit trivial $\Eqs$ and $\Neqs$ sets in configurations. We also omit the set enclosing brackets if the corresponding set is a singleton.

Let $C_0\seq\langle\underline{\pv\conc\pu_1\conc\cA\conc\pu_2\seq \pu_1\conc\cA\conc\pu_2\conc\pv},\OutTrue, \pv\neq \vz_1\conc\cA\conc\vz_2\rangle$ be the initial configuration. Then it may be unfolded either to $C_1\seq\langle \underline{\pu_1\conc\cA\conc\pu_2\seq\cA\conc\pu_2\conc\pu_1},\OutTrue,\pu_1\neq\vz_1\conc\cA\conc\vz_2)$ by the narrowing $\pv\rar\pu_1$ or to $C_2\seq\langle \underline{\pv\conc\pt\conc\pu'_1\conc\cA\conc\pu_2\seq\pt\conc\pu'_1\conc\cA\conc\pu_2\conc\pv},\OutTrue,\pv\neq\vz_1\conc\cA\conc\vz_2)$ by the narrowing $\pu_1\rar\pv\conc\pt\conc\pu'_1$. The narrowing $\pv\rar\pu_1\conc\cA\conc\pv'$ also satisfies the properties given in Definition~\ref{Definition:EquationTransition}, but leads to the contradiction with $\pv\neq\vz_1\conc\cA\conc\vz_2$.

After a number of applications of \textbf{UNFOLD} and \textbf{CLOSE}, the process tree of the equation is constructed. The tree is shown in the diagram below. 
 
$$
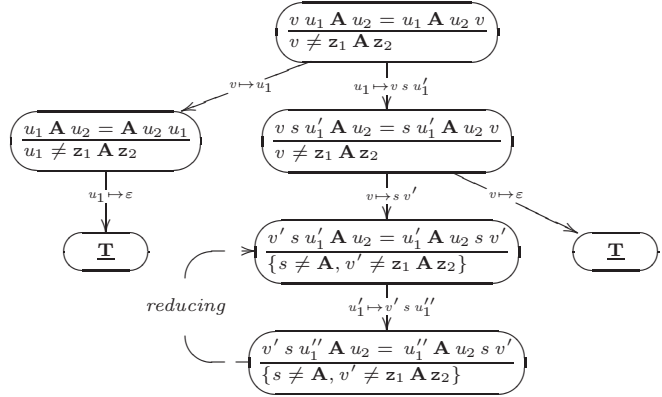
\scriptsize
\xymatrix @-2mm {
&\ov{\begin{array}{l} \underline{\pv\conc\pu_1\conc\cA\conc\pu_2\seq\pu_1\conc\cA\conc\pu_2\conc\pv} \\
\pv\neq\vz_1\conc\cA\conc\vz_2
\end{array}}\ar[dl]|{\skipper\pv\rar\pu_1}\ar[d]|{\skipper\pu_1\rar\pv\conc\pt\conc\pu'_1}
\\
\ov{\begin{array}{l}\underline{\pu_1\conc\cA\conc\pu_2\seq\cA\conc\pu_2\conc\pu_1}\\
\pu_1\neq\vz_1\conc\cA\conc\vz_2 \end{array}}\ar[d]|{\skipper\pu_1\rar\empt}&
\ov{\begin{array}{l}\underline{\pv\conc\pt\conc\pu'_1\conc\cA\conc\pu_2\seq\pt\conc\pu'_1\conc\cA\conc\pu_2\conc\pv}
\\\pv\neq\vz_1\conc\cA\conc\vz_2 \end{array}}\ar[d]|{\skipper\pv\rar\pt\conc\pv'}\ar[dr]|{\skipper\pv\rar\empt}\\
\ov{\quad\begin{array}{l}\underline{\OutTrue}\end{array}\quad}&
\ov{\begin{array}{l}\underline{\pv'\conc\pt\conc\pu'_1\conc\cA\conc\pu_2\seq\pu'_1\conc\cA\conc\pu_2\conc\pt\conc\pv'}\\
\{\pt\neq\cA, \pv'\neq\vz_1\conc\cA\conc\vz_2\}
\end{array}}\ar[d]|{\skipper\pu'_1\rar\pv'\conc\pt\conc\pu''_1}
&
\ov{\quad\begin{array}{l}\underline{\OutTrue}\end{array}\quad}
\\
&
\ov{\begin{array}{l}\underline{\pv'\conc\pt\conc\pu''_1\conc\cA\conc\pu_2\seq\conc\pu''_1\conc\cA\conc\pu_2\conc\pt\conc\pv'}\\
\{\pt\neq\cA, \pv'\neq\vz_1\conc\cA\conc\vz_2\}
\end{array}}\ar@{-->}`l[ul]-<-30pt,13pt>`[ur]|{\txt{reducing}}[u]
}
$$

All leaves in the subtree starting from the narrowing $\pv\rar\pt\conc\pv'$ are closed with elements of this subtree. Hence, the subtree presents an unreachable set of paths of the equation solving algorithm and the \textbf{PRUNE} rule should be used. The result is shown on the following diagram. A simple analysis of the diagram shows that for every solution of the equation $\pv\conc\pu_1\conc\cA\conc\pu_2\seq\pu_1\conc\cA\conc\pu_2\conc\pv$ provided that $\pv\neq\vz_1\conc\cA\conc\vz_2$, the property $\pv\seq\empt$ holds.

$$
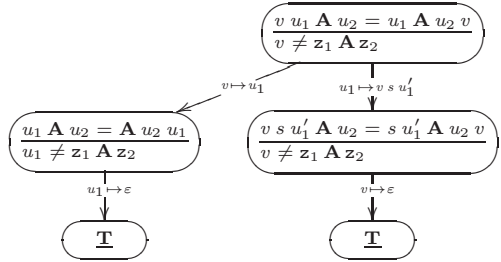
\scriptsize
\xymatrix @-2mm {
&\ov{\begin{array}{l} \underline{\pv\conc\pu_1\conc\cA\conc\pu_2\seq\pu_1\conc\cA\conc\pu_2\conc\pv} \\
\pv\neq\vz_1\conc\cA\conc\vz_2
\end{array}}\ar[dl]|{\skipper\pv\rar\pu_1}\ar[d]|{\skipper\pu_1\rar\pv\conc\pt\conc\pu'_1}
\\
\ov{\begin{array}{l}\underline{\pu_1\conc\cA\conc\pu_2\seq\cA\conc\pu_2\conc\pu_1}\\
\pu_1\neq\vz_1\conc\cA\conc\vz_2 \end{array}}\ar[d]|{\skipper\pu_1\rar\empt}&
\ov{\begin{array}{l}\underline{\pv\conc\pt\conc\pu'_1\conc\cA\conc\pu_2\seq\pt\conc\pu'_1\conc\cA\conc\pu_2\conc\pv}
\\\pv\neq\vz_1\conc\cA\conc\vz_2 \end{array}}\ar[d]|{\skipper\pv\rar\empt}\\
\ov{\quad\begin{array}{l}\underline{\OutTrue}\end{array}\quad}&
\ov{\quad\begin{array}{l}\underline{\OutTrue}\end{array}\quad}
}$$
\end{Example}

\begin{Example}\label{Example:NarrEqNarrNeq}

The function $\FPref$ below tests whether its second argument is a prefix of the first one.
$$
\xymatrix@-4.5mm {
*+[F-]{\begin{array}{llll}
\Rule^{\FPref}_1: & \FPref(\vx,\vx)&=&\OutTrue;\\
\Rule^{\FPref}_2: & \FPref(\empt,\vy)&=&\OutFalse;\\
\Rule^{\FPref}_3: & \FPref(\vx\conc\vs,\vy)&=&\FPref(\vx,\vy); 
\end{array}
}}
$$

What configurations $C_i\seq\langle \Phi_i,\Eqs_i,\Neqs_i\rangle$ can be generated from $$C_0\seq\langle \FPref(\pu\conc\cA\conc\pv,\cA\conc\pu),\pu\conc\pv\seq\pv\conc\pu,\pv\neq \vz_1\conc\cA\conc\vz_2\rangle$$ by \textbf{UNFOLD} rule? Below we comment the answer to this question and demonstrate it with the picture.

Consider the rule $\Rule^{\FPref}_1$. It produces the only pair of the narrowing substitution $\Narr_1\colon\pv\rar\empt$ and the narrowing equation $E\colon \pu\conc\cA\seq\cA\conc\pu$. Hence, $\Rule^{\FPref}_1$ produces the only configuration $C_1\seq\langle \OutTrue, \OutTrue,\OutTrue\rangle$.

$C_0$ cannot be matched against $\Rule^{\FPref}_2$, because  the algorithm solving the equation $\pu\conc\cA\conc\pv\seq\empt$ results in the contradiction. 

The rule $\Rule^{\FPref}_3$ generates the two narrowing substitutions: $\Narr_2\colon \pv\rar\pv'\conc\pt$,  $\Narr_3\colon\pv\rar\empt$. No narrowing equations are generated. Let us denote the configurations generated by $\Narr_2$ and $\Narr_3$ as $C_2\seq \langle \Phi_2,\Eqs_2,\Neqs_2\rangle$ and $C_3\seq\langle\Phi_3,\Eqs_3,\Neqs_3\rangle$, respectively.

$\Phi_2\seq\FPref(\pu\conc\cA\conc\pv',\cA\conc\pu)$. $\Eqs_2$ contains the only equation $\pu\conc\pv'\conc\pt\seq\pv'\conc\pt\conc\pu$. The inequality $\pv'\conc\pt\neq\vz_1\conc\cA\conc\vz_2$ implies $\pv'\neq\vz_1\conc\cA\conc\vz_2$ and $\pt\neq\cA$, which are included in $\Neqs_2$. The set of the narrowing inequalities $\Neqs_{Scr}$ is trivial, because $\Rule^{\FPref}_1$ is applied only if $\pv\seq\empt$, while $\pv\Narr_2\neq\empt$.  

$\Phi_3\seq\FPref(\pu,\cA\conc\pu)$. $\Eqs_3$ contains the only trivial equation $\pu\seq\pu$, replaced by $\OutTrue$. There are no descendants of inequality $\pv\neq\vz_1\conc\cA\vz_2$. Thus, the set $\Neqs_{Subst}$ of $C_3$ is trivial. Set $\Neqs_{Scr}$ must contain corollaries from the inequality $\pu\conc\cA\neq\cA\conc\pu$ corresponding to the narrowing equation  of rule $\Rule^{\FPref}_1$, but this inequality is not linear and cannot be decomposed to linear inequalities, hence $\Neqs_{Scr}$ is also trivial.

$$
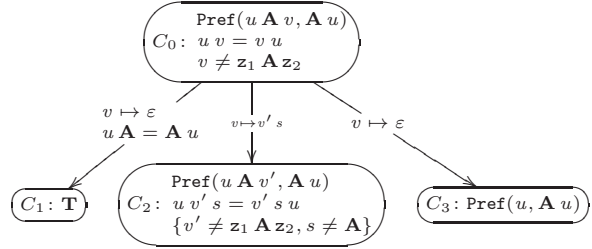
\scriptsize
\xymatrix @-4mm {
&\ov{C_0\colon \begin{array}{l}\FPref(\pu\conc\cA\conc\pv,\cA\conc\pu) \\ 
\pu\conc\pv\seq\pv\conc\pu \\
\pv\neq\vz_1\conc\cA\conc\vz_2\end{array}}\ar[ddl]|{\begin{array}{l}\pv\rar\empt\\ \pu\conc\cA\seq\cA\conc\pu\end{array}}\ar|{\skipper\pv\rar\pv'\conc\pt}[dd]
\ar|{\begin{array}{l}\pv\rar\empt \end{array}}[ddr]\\\\
\ov{C_1\colon \OutTrue}&
\ov{C_2\colon \begin{array}{l} \FPref(\pu\conc\cA\conc\pv',\cA\conc\pu)\\
\pu\conc\pv'\conc\pt\seq\pv'\conc\pt\conc\pu\\
\{\pv'\neq\vz_1\conc\cA\conc\vz_2, \pt\neq\cA\}\end{array}}&\ov{C_3\colon \FPref(\pu,\cA\conc\pu)}
}
$$

The branches in the process tree are ordered: only if the condition on the branch leading to $C_1$ fails, a computation may follow the branch $C_3$. But we do not use this order for our reachability analysis.   

\end{Example}

\begin{Example}\label{Example:irreg}

The analysis below uses the expressiveness of the language of the word equations. A constraint, which is not expressible by a context-free grammar, is introduced and used for the analysis.

We consider the following program.
$$
\xymatrix@-6mm {
*+[F-]{\begin{array}{lll}
\Go(\vx,\vy)&=&\BothA(\Iter(\vx,\vy,\empt),\vx)\\
\vspace{-8pt}\\
\Iter(\vx,\empt,\vy)&=&\vy;\\
\Iter(\vx,\vs\conc\vy,\vz)&=&\Iter(\vx,\vy,\vz\conc\vx); \\
\vspace{-8pt}\\
\BothA(\vx_1\conc\cA\conc\vx_2,\vy_1\conc\cA\conc\vy_2)&=&\OutTrue; \\
\BothA(\vx_1\conc\cA\conc\vx_2,\vy)&=&\OutFalse; \\
\BothA(\vx,\vy_1\conc\cA\conc\vy_2)&=&\OutFalse; \\
\BothA(\vx,\vy)&=&\OutTrue; \\
\end{array}}}
$$
The second argument of $\Iter$ serves as a unary Peano number which determines the number of concatenations of $\vx$ to the third argument.

The function $\BothA$ checks whether both of its arguments contain $\cA$. If both or none of them contain $\cA$, the value $\OutTrue$ is returned. Otherwise, $\OutFalse$ is returned. 

Let us apply our analysis scheme to the computation starting at the input point $\Go(\pu,\pv)\seq\BothA(\Iter(\pu,\pv,\empt),\pu)$. The initial predicates set is trivial. 
The result of the first application of \textbf{UNFOLD} to $\langle \BothA(\Iter(\pu,\pv,\empt),\pu),\OutTrue,\OutTrue \rangle$ is shown in the following scheme.
$$
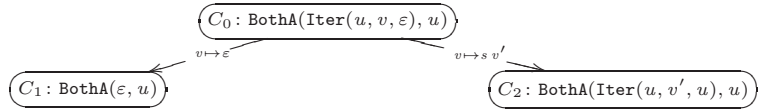
\scriptsize
\xymatrix @-4mm {
&\ov{C_0\colon\BothA(\Iter(\pu,\pv,\empt),\pu)}\ar[dl]|{\skipper\pv\rar\empt}\ar[dr]|{\skipper\pv\rar\pt\conc\pv'}\\
\ov{C_1\colon\BothA(\empt,\pu)}&&\ov{C_2\colon\BothA(\Iter(\pu,\pv',\pu),\pu)}
}$$

When \textbf{UNFOLD} is applied to $C_1$, two children nodes with the constant configurations are generated. The third rule of $\BothA$ definition generates narrowing $\pu\rar\pu_1\conc\cA\conc\pu_2$ and the node labeled with $\langle\OutFalse, \OutTrue, \OutTrue\rangle$. The fourth rule of $\BothA$ definition generates the node labeled with $\langle \OutTrue,\OutTrue,\OutTrue\rangle$: the set $\Neqs_{Scr}$ corresponding to the configuration contains inequality $\pu\neq\vz_1\conc\cA\conc\vz_2$, but the parameter $\pu$ is used in no more equations and expressions of the configuration, thus, the inequality is omitted. No contradiction is generated in $C_1$.

After the algorithm checks that $C_2$ does not generate the contradiction (by applying \textbf{UNFOLD} to $C_2$), \textbf{GENERALIZE} is activated and $C_0$ is generalized with $C_2$. Expressions $\empt$ and $\pu$ are generalized to a new parameter $\pw_2$, which is supplied by the additional equation $\pu\conc\pw_2\seq\pw_2\conc\pu$ (see \sectionname~\ref{subsection:genrules}). This equation expresses the following non-regular property of $\pu$ and $\pw_2$: $\exists q,n,m(q\in\Sigma^*\logand \pu\seq q^n\logand \pw_2\seq q^m)$. There $q^n$ stands for the word $\underbrace{q\conc q\dots\conc q}_n$. We have to throw out the previous calculations and unfold the generalized configuration. 

After several steps of the scheme application, the following process tree is built.

$$
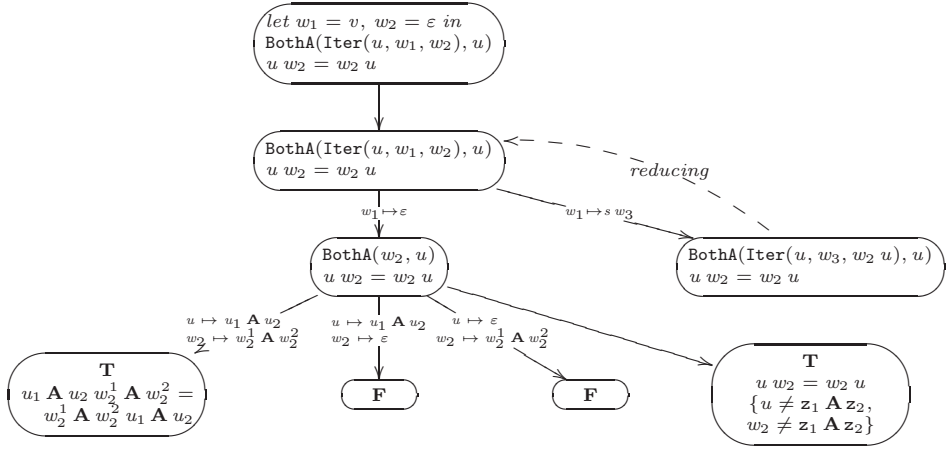
\scriptsize
\xymatrix @-2mm {
&\ov{\begin{array}{l}\tlet \pw_1=\pv,\,\pw_2=\empt \tin \\
\BothA(\Iter(\pu,\pw_1,\pw_2),\pu) \\
\pu\conc\pw_2\seq\pw_2\conc\pu
\end{array}}\ar[d]\\
&\ov{\begin{array}{l}\BothA(\Iter(\pu,\pw_1,\pw_2),\pu)\\\pu\conc\pw_2\seq\pw_2\conc\pu
\end{array}}\ar[d]|{\skipper\pw_1\rar\empt}\ar[drr]|{\skipper\pw_1\rar\pt\conc \pw_3}\\
&\ov{\begin{array}{l}
\BothA(\pw_2,\pu) \\
\pu\conc\pw_2\seq\pw_2\conc\pu
\end{array}}\ar[dl]|{\tiny\begin{array}{l}\pu\rar\pu_1\conc\cA\conc\pu_2 \\ \pw_2\rar\pw^1_2\conc\cA\conc\pw^2_2 \end{array}}
\ar[d]|{\tiny\begin{array}{l}\pu\rar\pu_1\conc\cA\conc\pu_2 \\  \pw_2\rar\empt \end{array}}
\ar[dr]|{\tiny\begin{array}{l}\quad\quad\pu\rar\empt \\\quad\pw_2\rar\pw^1_2\conc\cA\conc\pw^2_2 \end{array}}
\ar[drr]
&&\ov{\begin{array}{l}\BothA(\Iter(\pu,\pw_3,\pw_2\conc\pu),\pu)\\ 
\pu\conc\pw_2\seq\pw_2\conc\pu\end{array}}\ar@{-->}@/_2pc/|<(.3){\txt{reducing}}[llu]
\\
\ov{\begin{array}{c}
\OutTrue\\
\pu_1\conc\cA\conc\pu_2\conc\pw^1_2\conc\cA\conc\pw^2_2\seq \\
\quad\pw^1_2\conc\cA\conc\pw^2_2\conc\pu_1\conc\cA\conc\pu_2
\end{array}}
&
\ov{\quad\OutFalse\quad}
&
\ov{\quad\OutFalse\quad}
&
\ov{\quad\begin{array}{c}
\OutTrue\\
\pu\conc\pw_2\seq \pw_2\conc\pu\\
\{\pu\neq\vz_1\conc\cA\conc\vz_2,\\
\pw_2\neq\vz_1\conc\cA\conc\vz_2\}
\end{array}\quad}
}$$

The narrowings leading to $\OutFalse$ nodes are generated by the constraint analysis given in Example~\ref{Example:EquationSolving}. The branch labeled with the narrowing $\pu\rar\empt,\,\pw_2\rar\pw^1_2\conc\cA\conc\pw^2_2$ cannot be pruned, because the generalization carries no information contradicting these narrowings. 

If $C_0$ and $C_2$ are generalized using the disjunction generation (see the second rule for equations generated by generalizations, \sectionname~\ref{subsection:genrules}), then the following equations are introduced. $\pv_1$, $\pv_2$, $\pv_3$, $\pv_4$ are all fresh parameters.

$$
\begin{array}{l}
\pv_1\conc ()\conc\pw_2\conc()\conc\pv_2\seq  ()\conc()\conc\pv_3\conc\pu\conc\pv_4\conc ()\\
\pu\conc\pw_2\seq\pw_2\conc\pu
\end{array}
$$   

Based on conjunction of these predicates over $\pw_2$ and $\pu$, the algorithm described above is able to prove that the branch with the narrowing $\pu\rar\empt,\,\pw_2\rar\pw^1_2\conc\cA\conc\pw^2_2$ is unreachable. A simple analysis of the resulting process tree shows that the call $\Go(\pu,\pv)$ can compute $\OutFalse$ only if $\pv\seq\empt$.

\end{Example}

\begin{Example}\label{Example:Sweden}
The program below is taken from \cite{Abdulla}. The verification task is to check when it can return $\OutTrue$. Using our scheme, a supercompiler easily proves that call $\Go(\pu,\pv,\pw)$ returns $\OutTrue$ only if $\pw\seq\empt$. Note that the function $\FGram$ is a generating context-free grammar.

$$
\xymatrix@-4.5mm {
*+[F-]{\begin{array}{llll}
\Go(\vx,\vy,\vz)&=&\FCheck(\vx,\vy,\FGram(\vz));\\
\\
\FGram(\cI\conc \vx) & = &\cA\conc\FGram(\vx)\conc \cB;\\
\FGram(\cS\conc \vx) & = &\FGram(\vx)\conc \cB;\\
\FGram(\empt) & = &\empt;\\
\\
\FCheck(\vx,\vy,\vz)&=&\FEqual(\cA\conc\vz\conc\vx,\vz\conc\vy);\\
\FEqual(\vx,\vx)&=&\OutTrue;\\
\FEqual(\vx,\vy)&=&\OutFalse;
\end{array}}}
$$
\end{Example}

\end{document}